\documentclass{aa}
\usepackage{graphicx}
\newcommand{\HI}{{\ion{H}{i}}}
\begin{document}
\title{Constraining the spectral age of very asymmetric CSOs:}
\subtitle{Evidence of the influence of the ambient medium }
\author{
M. Orienti\inst{1,2} \and
D. Dallacasa\inst{1,2}\and 
C. Stanghellini\inst{2}
}
\offprints{M. Orienti}
\institute{
Dipartimento di Astronomia, Universit\`a di Bologna, via Ranzani 1,
I-40127, Bologna, Italy \and 
Istituto di Radioastronomia -- INAF, via Gobetti 101, I-40129, Bologna,
Italy 
}
\date{Received \today; accepted ?}

\abstract
{}
{We constrain the spectral ages for two very asymmetric Compact
  Symmetric Objects (CSO) from the B3-VLA-CSS sample, and we
  investigate the role of the ambient medium potentially 
able to influence the
  individual source evolution.}
{ Multi-frequency VLBA observations have been carried out to study the
  distribution of the break frequency of the spectra across different
  regions of each source.} 
{ From the analysis of synchrotron spectra and assuming an equipartition
  magnetic field, we find radiative ages of about 2$\times$10$^{3}$
  and 10$^{4}$ years for B0147+400 and B0840+424, 
  respectively. 
  The derived individual hot-spot advance speed is in the range between
  0.03c and 0.3c, in agreement with kinematic studies carried out on
  other CSOs. The very asymmetric morphology found in both sources
  is likely related to an inhomogeneous ambient medium in which the
  sources are growing, rather than to different intrinsic hot-spot
  pressures on the two sides.}
{}
\titlerunning{Constraining the spectral age of very asymmetric CSOs}
\maketitle
\section{Introduction}
Powerful and intrinsically compact ($\leq$ 1'') radio sources with
convex radio spectra peaking between 100 MHz and a few GHz represent
a significant fraction ($\sim$ 15\%) of the objects in 
flux density-limited radio source catalogues.   
When imaged with parsec-scale resolution, they often display a
symmetric radio structure dominated by hot-spots and mini-lobes,
namely a scaled-down version (from 0.1 to a few kpc) of powerful,
edge-brightened radio galaxies known as classical doubles. These radio
sources, termed ``Compact Symmetric Objects'' (CSOs) by Wilkinson et
al. (\cite{wil94}), are currently considered the early stages of the
large radio sources (Fanti et al. \cite{cf95}; Readhead et
al. \cite{rh96}; Snellen et al. \cite{sn00}). \\
Two independent pieces of evidence strongly support the {\it youth
scenario}: the kinematic study of hot-spot separation velocities
(e.g. Polatidis \& Conway \cite{pc03}; Gugliucci et al. \cite{gu05}),
and the measure of radiative age from the analysis of the source 
spectrum (Murgia \cite{mm03}; Murgia et al. \cite{mm99}); both
find ages of about 10$^{3}$ -- 10$^{5}$ years.\\
On the other hand, the alternative model, the {\it frustration
scenario} (Baum et al. \cite{baum90}; van Breugel et
al. \cite{breu84}), which postulates that the radio source is
trapped by an unusually dense gas, lacks any observational
evidence, at least for the CSOs observed so far
(see e.g. Fanti et al. \cite{cf00}; Siemiginowska et
al. \cite{sa05}).\\    

Synchrotron spectral ageing is based on the determination of the
spectral break, which occurs at progressively lower frequencies as time
passes. Indeed, according to various source growth models
(i.e. Kardashev \cite{karda62}; Pacholczyk \cite{pacho70}; Jaffe \&
Perola \cite{jp74}), relativistic electrons located in different
source regions have been deposited at different times, and
locally the electron age measures the time elapsed since their
production and/or latest acceleration ($t_{\rm syn}$) when they
crossed the hot-spot during its outward motion. The radiative age can
be easily computed once the break frequency ($\nu_{\rm br}$) and the
magnetic field ($B$) are known:\\ 

\begin{equation}
t_{\rm syn} \propto B^{-3/2} \nu_{\rm br}^{-1/2}.
\label{equation1}
\end{equation}

So far, most if not all the works on the measure of the source
radiative age are based on the spectral break derived from the 
source-integrated spectra.  
One disadvantage of this approach is that the contributions of the various
source components (core, jets, hot-spots and lobes), each one with its
own spectral shape, are all mixed together. 
The brightest component is the one that influences the age
determination the most; if we
consider a source whose emission is dominated by the hot-spots where
electrons are likely to be re-accelerated, the spectral ages derived
can be completely unrelated to the source age.\\ 
On the other hand, the radiative age of the back-flow tail in the lobes
measures the time elapsed since the last acceleration of those particles,
namely when the hot-spot crossed that location during its outward
expansion. The best (i.e. oldest) measure of the radiative age then
comes from the innermost edges of the lobes, where the electrons 
were deposited at the very beginning. \\
Multi-frequency NRAO VLBA images can be very effective in
constraining the
source age in small radio sources whose radio emission contains
a substantial contribution from the mini-lobes. 
Using multi-frequency images with pc-scale resolution
it is possible to study the spectral ageing in intrinsically very
small radio galaxies: it is important to distinguish regions in which
electrons are injected/accelerated (i.e. core and hot-spots) from
those in which electrons age, like in
the back-flow tails of the mini radio lobes.\\ 
 
This paper reports the results of new multi-frequency 
L (1.4-1.6 GHz), C (4.5-4.9 GHz) and X (8.1-8.5 GHz)
bands) VLBA observations of two CSOs from the B3-VLA-CSS sample 
(Fanti et al. \cite{cf01}): B0147+400 and B0840+424. For both
sources the core has been clearly detected. 

Throughout this paper, we assume H$_{0}$= 71 km s$^{-1}$ Mpc$^{-1}$,
$\Omega_{\rm M}$ = 0.27 and $\Omega_{\rm \lambda}$ = 0.73, in a flat
Universe. \\

\begin{table*}
\begin{center}
\begin{tabular}{cccccccccc}
\hline
\hline
&&&&&&&&&\\
Source&$\nu$&Obs. date&Missing&$u_{\it min}$&$u_{\it max}$& &Beam& &Noise\\
B1950&GHz& &antennas&M$\lambda$&M$\lambda$ &mas&mas&$\circ$&mJy/beam\\
(1)&(2)&(3)&(4)&(5)&(6)&(7)&(8)&(9)&(10)\\
\hline
&&&&&&&&&\\
B0147+400&1.407&09/06/2004&  &0.16&41&9.3&5.3&-3&0.21\\
         &1.643&          &  &0.18&50&8.3&4.7&-4&0.38\\
         &4.543&          &  &0.49&110&2.9&1.7&-7&0.11\\
         &4.893&          &  &0.50&140&2.7&1.6&-5&0.10\\
         &8.115&          &  &0.99&230&1.7&1.0& 5&0.08\\
         &8.493&          &  &1.05&250&1.6&0.9& 5&0.07\\
B0840+424&1.407&11/24/2004&HN&0.15&40&12.0&7.5&-2&0.19\\
         &1.643&          &HN&0.18&46&10.5&6.9& 0&0.17\\
         &4.543&          &SC&0.48&110&3.5&1.9&-20&0.15\\
         &4.893&          &SC&0.50&125&3.4&1.8&-20&0.14\\
         &8.115&          &SC&0.83&210&2.0&1.6&-23&0.16\\
         &8.493&          &SC&0.88&230&2.0&1.5&-18&0.12\\
&&&&&&&&&\\
\hline
\end{tabular} 
\vspace{0.5cm}
\end{center}   
\caption{Basic information on the VLBA observations: Column 1: Source
name (B1950); Col.2: Observed frequency; Col. 3: Date of observations;
Col. 4: missing antenna; Columns 5 and 6: shortest and longest projected
baselines; Columns 7, 8 and 9: beam major and minor axis and p.a. of
the full resolution images; Col. 10: 1$\sigma$ rms noise level on the
full resolution image plane.}   
\label{taboss}
\end{table*}

\section{Observations and data reduction} 
Pc-scale resolution observations were carried out in two different
runs using the VLBA plus a single VLA antenna, on September 6th
(B0147+400) and November 24th 2004 (B0840+424), in full polarization
mode with a recording band-width of 16 MHz at 128 Mbps, for a total of
18 hours. Each source was observed for about 100, 150 and 210 minutes
in L, C and X bands respectively, spread into short scans at various
hour angles to improve the {\it uv} coverage.
In order to achieve a wider frequency coverage, in each observing band
we observed with two 8-MHz sub-bands 
widely spaced in frequency, obtaining
6 independent images, which improve the ability to
  constrain the spectral models. Details on the observations are
summarized in Table {\ref{taboss}}.\\
The correlation was performed at the VLBA correlator in Socorro, and
data reduction was carried out with the NRAO AIPS package. 
After the application of system temperature and antenna gain
information, the amplitudes were checked using the data on DA\,193
(J0555+398) which is unresolved on a large subset of baselines at all
frequencies, and whose flux density is monitored at the VLA in C and X
bands. \\
In both observations the sources DA\,193 and J0927+3902 were both used
to generate the bandpass correction.\\
The error on the absolute flux density scale can be estimated within 3
- 5\% on the basis of the fluctuations of the amplitude gain
solutions. \\ 
In C and X bands the instrumental polarization was removed by
using the AIPS task PCAL; the absolute orientation of the electric
vector of DA\,193 and J0927+3902 was compared with the VLA/VLBA
polarization calibration database to derive the corrections. The
values derived from the two sources were in excellent agreement ($\leq$
2$^{\circ}$).  The calibration of the instrumental polarization
was not performed for the L band data. \\
The images at each individual frequency were obtained after a number
of phase-only self-calibration iterations. Information on the full
resolution Stokes I images is given in Table \ref{taboss}. 
Stokes
U and Q images were produced from the final dataset.

As last step, we produced 
images at 1.6, 4.5, 4.9,
8.1 and 8.5 GHz, using natural grid weighting and the same {\it
uv-range} common to all the observing frequencies for each source
(1.05 - 41 M$\lambda$ and 0.88 - 40 M$\lambda$ for B0147+400 
and B0840+424 respectively), 
in order to have almost the same {\it uv} and image sampling,
as well as restoring beam, as for the 1.4 GHz data. \\

For each source, these {\it low-resolution} images were combined to
produce a multi-frequency data cube, which was then analyzed by the
synage++ software (Murgia \cite{mmphd}) 
for subsequent spectral studies. Image
registration was checked by comparing the location of optically thin
bright features. 
In this paper, we do not show the {\it full-resolution} images
since they do not add any new information 
to those presented by Dallacasa et
al. (\cite{dd02}) and Orienti et al. (\cite{mo04}).\\

\section{Multi-frequency spectral analysis}

Spectral index imaging is quite a hard task for VLBI experiments,
since it is difficult to obtain well matched {\it uv-}coverages at the
various observing frequencies. In particular there is a lack of 
short spacings at high frequencies. \\ 
In our VLBA observations, the key addition of a single VLA antenna makes 
the differences in the sampling density at short spacing less
effective, allowing us to produce high-resolution spectral index images.
Furthermore, the availability of 6 independent frequencies allows us
to determine the source age by fitting the spectra in each pixel, with
a good confidence level.\\ 
The multi-frequency high-resolution images provided by VLBA allow the
determination of the nature of the source components (core, jets, lobes and
hot-spots) and therefore, allow us
to choose the best model to fit the
observed spectral shapes. 
Indeed, although the radiative losses always imply a high-frequency
steepening, the local spectral shape is strongly related to the
evolution of the emission from an electron population and by the
possible presence of injection of fresh relativistic electrons. 
For example, in the hot-spots, we expect that the observed spectra are well
fitted by models predicting a continuous injection of fresh particles,
while lobes and extended features will be better fitted by
single-injection models (i.e. JP, Jaffe \& Perola 
\cite{jp74}; KP, Kardashev \cite{karda62}; Pacholczyk \cite{pacho70})     
where the radiative losses play an important role in modifying the
initial spectral shape. \\    
A previous work on the spectral ageing in two CSOs (Murgia
\cite{mm03}) has shown that the break frequency decreases if we move
from regions near the hot-spot toward those located at the inner edges
of the lobe, in the core direction. 
This is consistent with the dynamical scenario in which the electrons
deposited at the centre of the source are older than those found
closer to the hot-spot, which is what is expected if the source is expanding 
with time and the principal
site of particle acceleration is the hot-spot.\\
Therefore, to obtain a measure of the source age, it is possible
to investigate how the break frequency changes across the lobes, where
electrons age without any further substantial acceleration and,
locally, without any new supply of fresh particles.\\

\section{Multi-frequency images of CSOs: constraining the age from the
spectra}

As mentioned in Section 3, the pc-scale resolution achieved by the
VLBA enables us to study extremely compact source components, found
to be unresolved with other radio telescopes. 
We perform a detailed spectral ageing study of the CSO sources
B0147+400 and B0840+424 from the B3-VLA-CSS sample (Fanti et
al. \cite{cf01}). 
Both radio sources are characterized by a weak core and two very
asymmetric, both in arm-length ratio and brightness, 
well-resolved mini-lobes.
In B0147+400 
the two mini-lobes lie roughly in the East-West direction, and
are connected by an extended, steep-spectrum ($\alpha$ $>$ 2) bridge
visible in the L band only, 
while 
in B0840+424 the mini-lobes are deployed roughly in the North-South
direction. \\ 
In the standard source model, the brightest component is also the
farthest from the core, since the differences in brightness and
arm-length ratio are due to beaming effects and path delay. 
In radio galaxies where the radio axis is oriented 
at large angles to the line
of sight, such asymmetries are expected to be quite small.
On the contrary,
in the two sources in this paper, the
brightest lobe is the one closest to the core, while the faintest one is
much further away, suggesting a strong influence exerted by the
ambient medium, which can be quite complex and inhomogeneous on such
small scales.
For a more detailed description of the morphologies and the physical
parameters of both sources, see Orienti et al. (\cite{mo04}) and
Dallacasa et al. (\cite{dd02}).\\ 

\subsection{The hot-spots}
\begin{figure*}
\begin{center}
\includegraphics{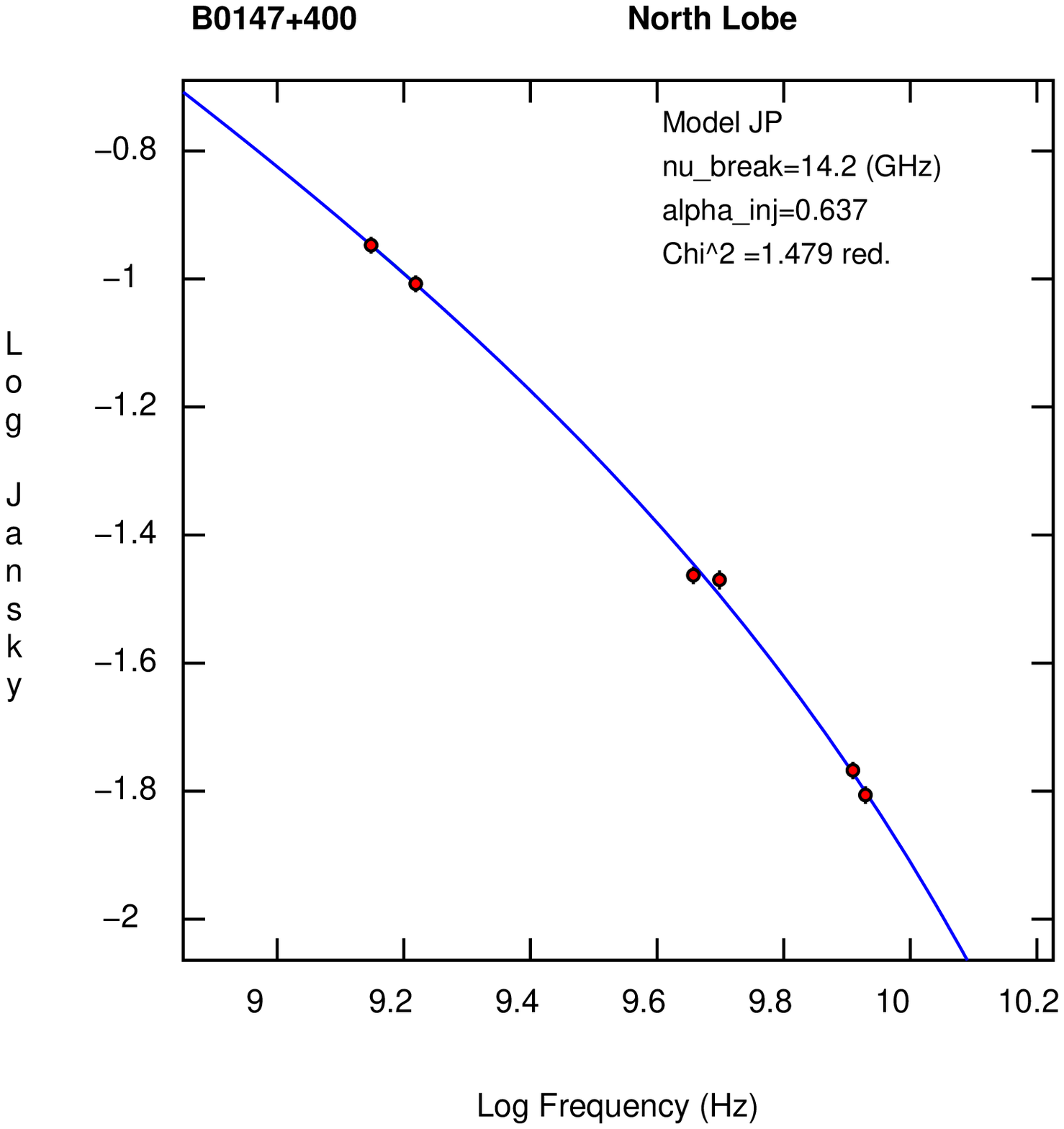}
\includegraphics{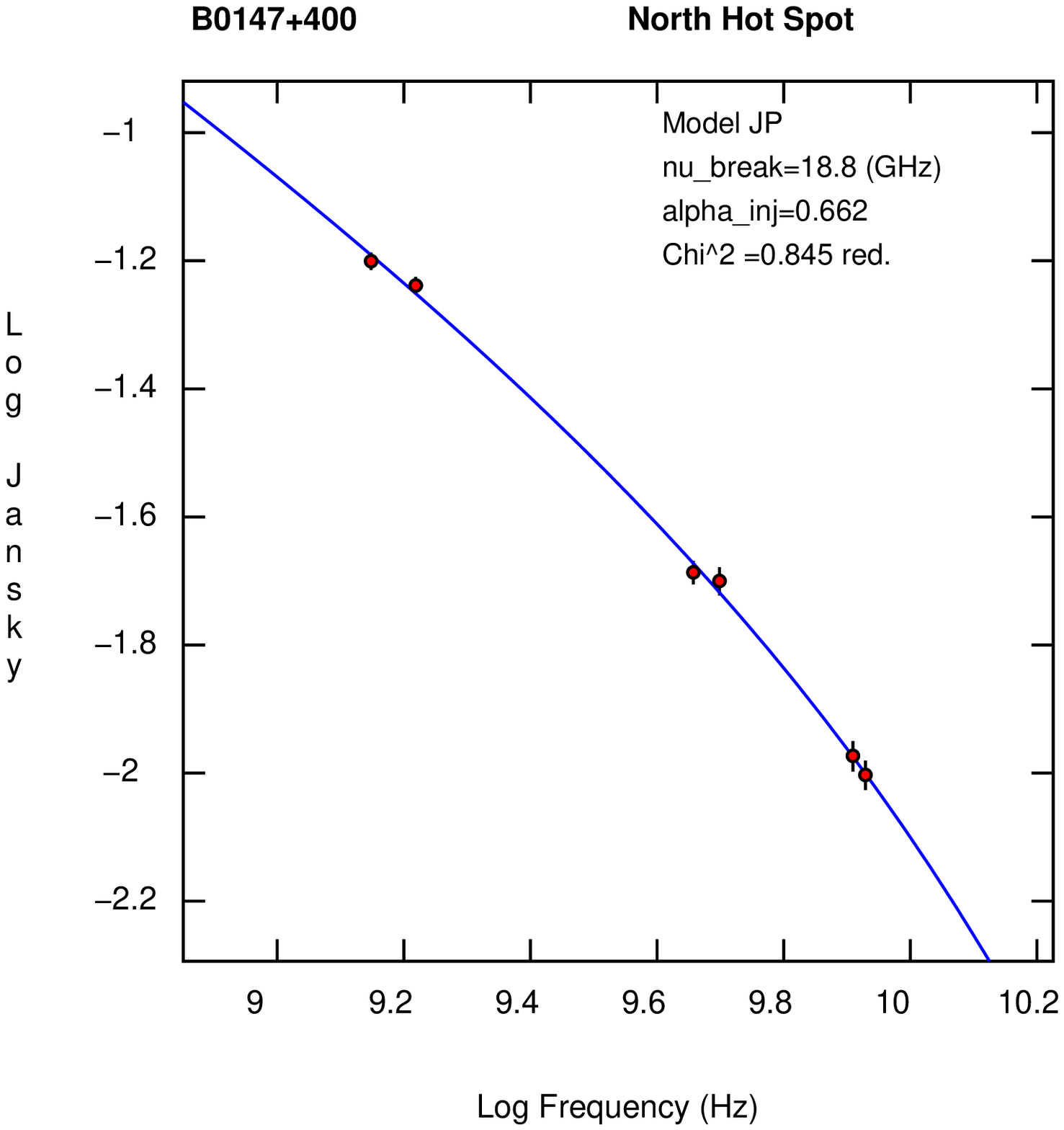}
\includegraphics{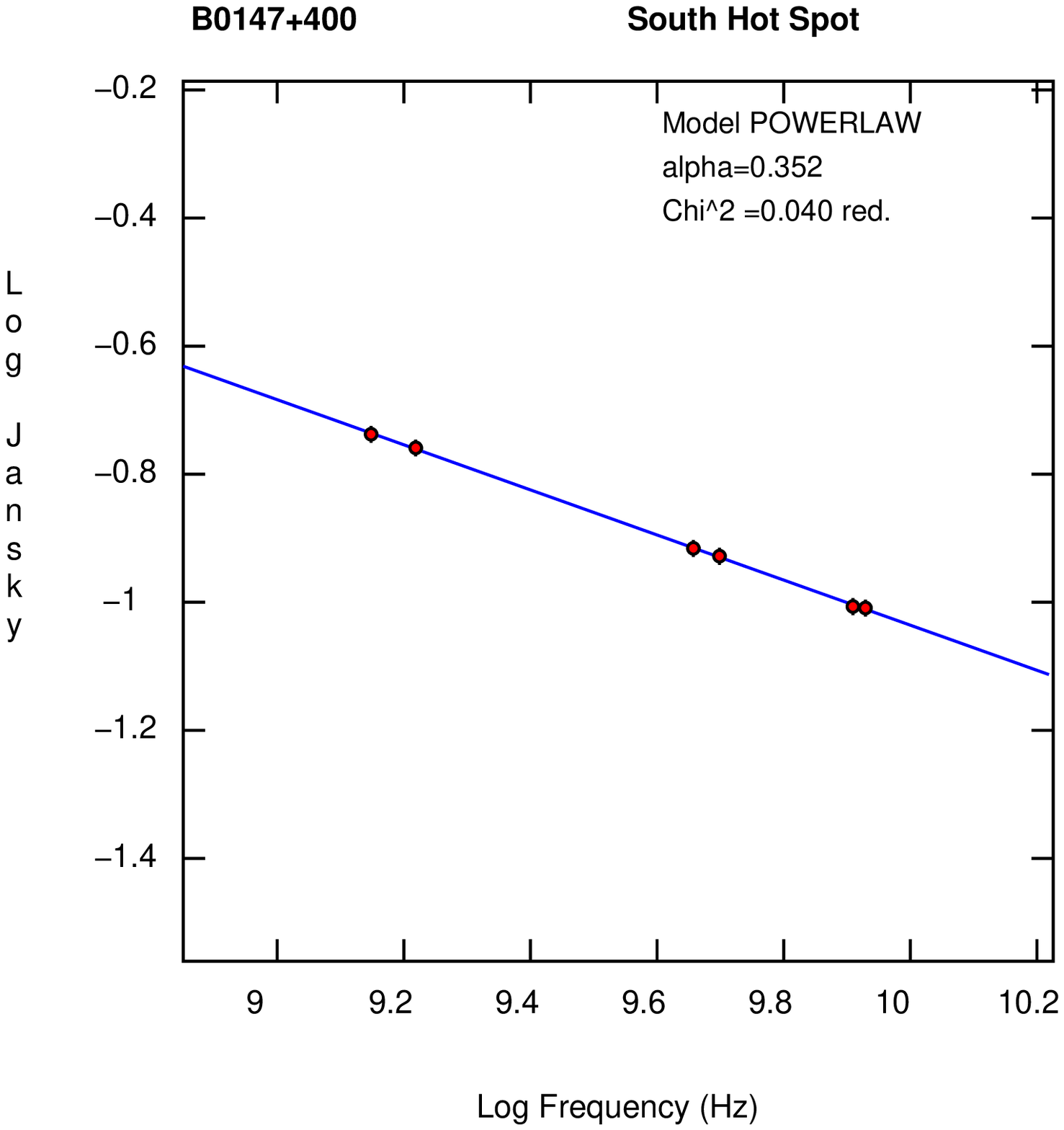}
\includegraphics{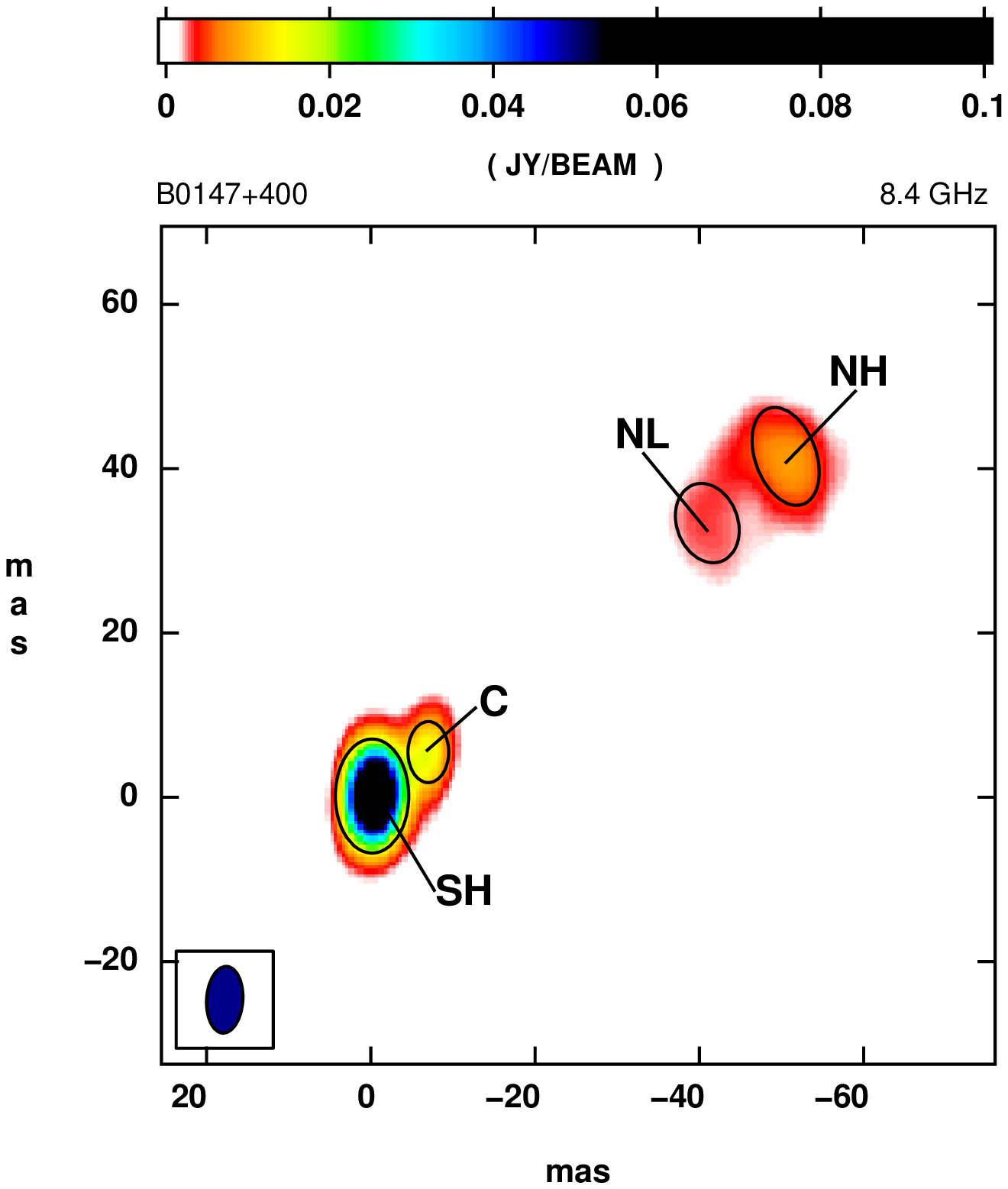}
\includegraphics{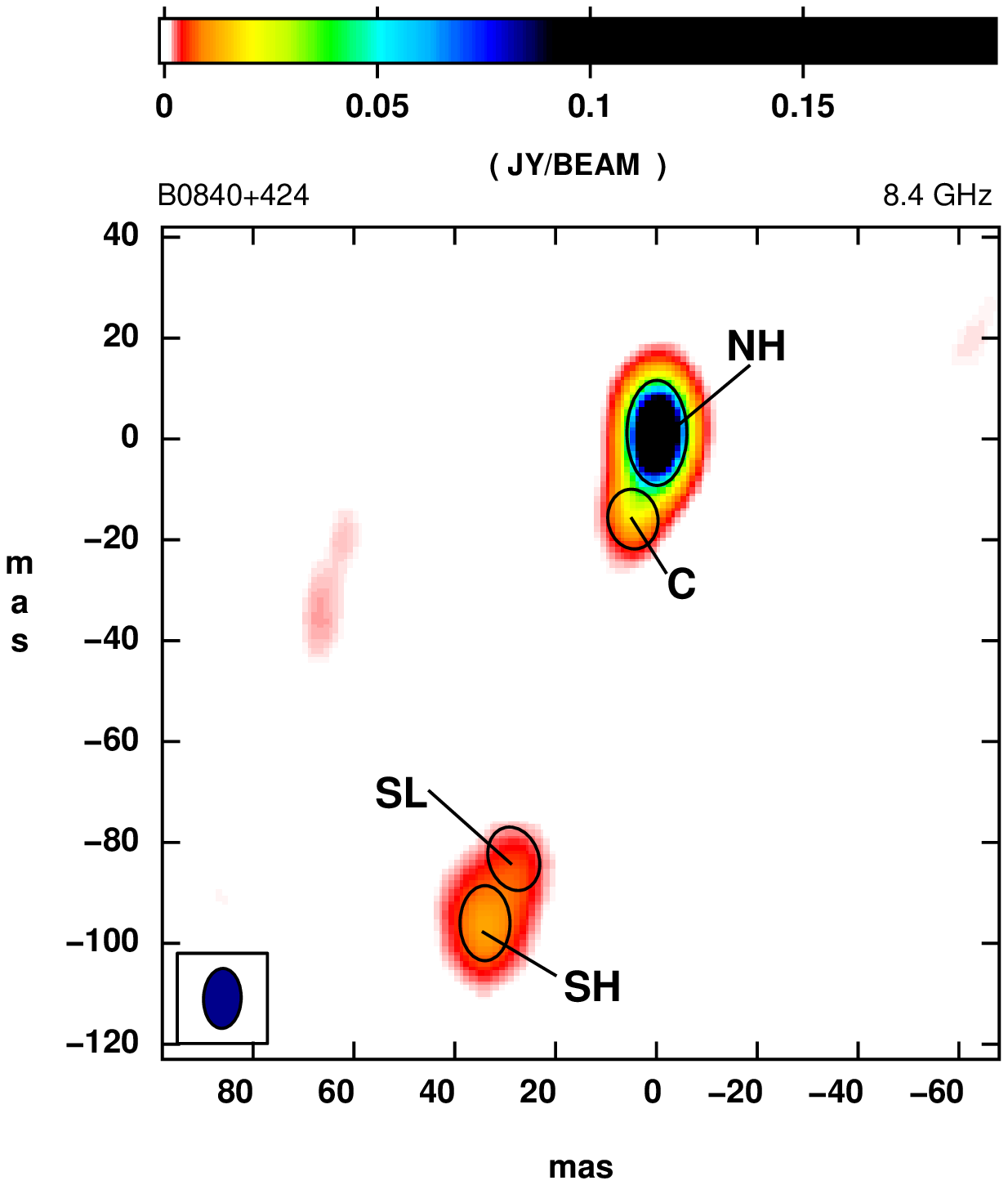}
\includegraphics{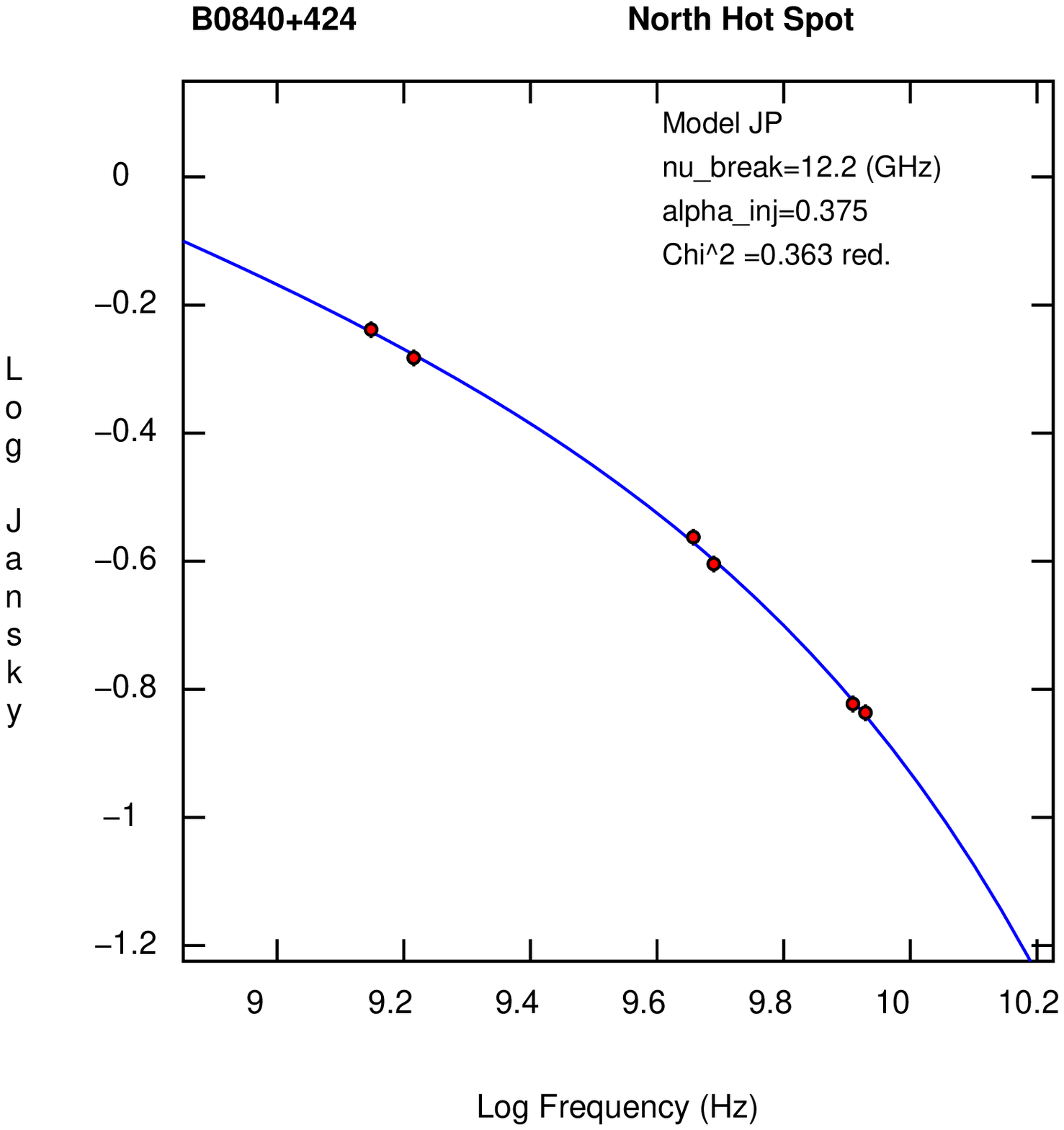}
\includegraphics{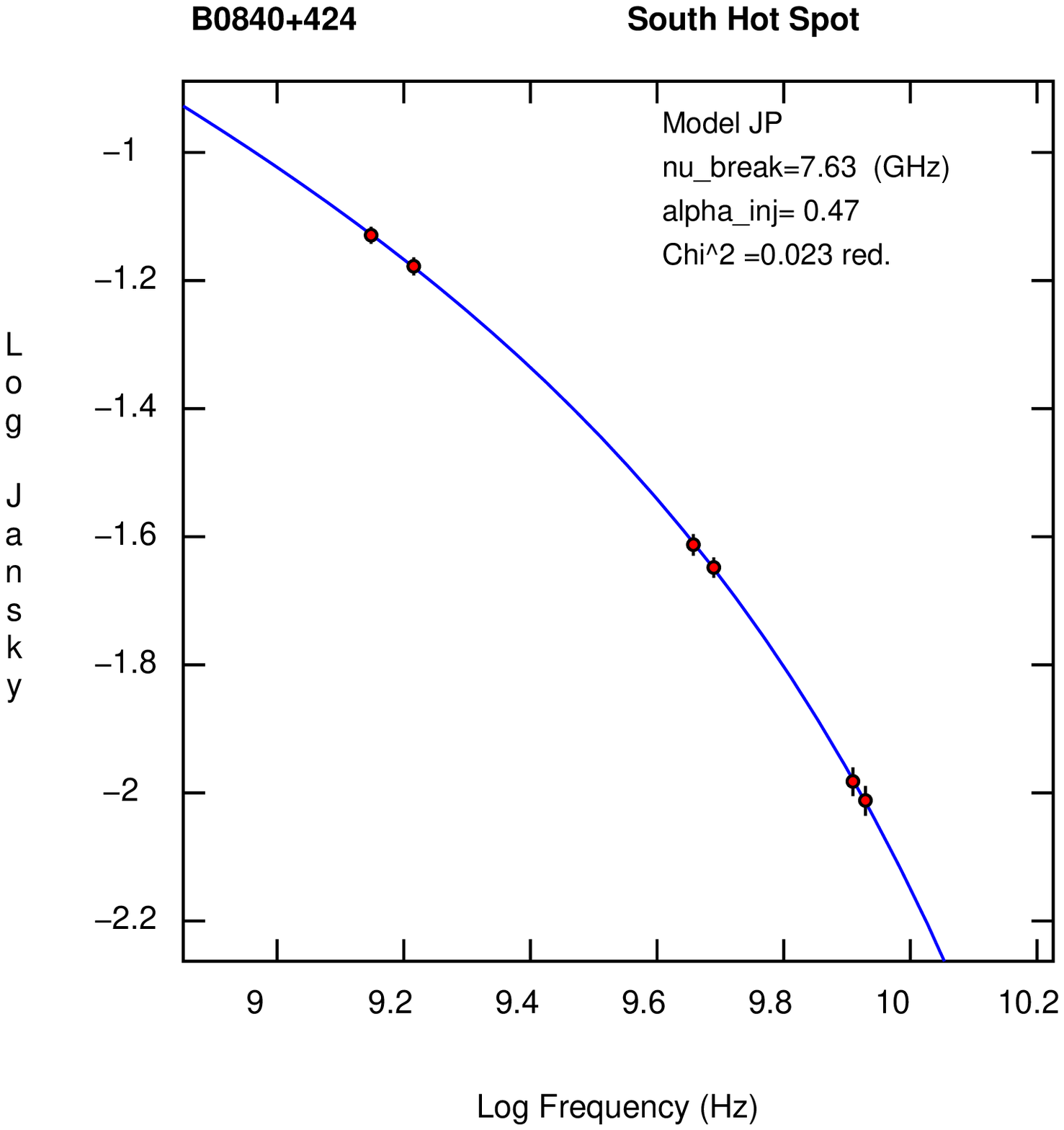}
\includegraphics{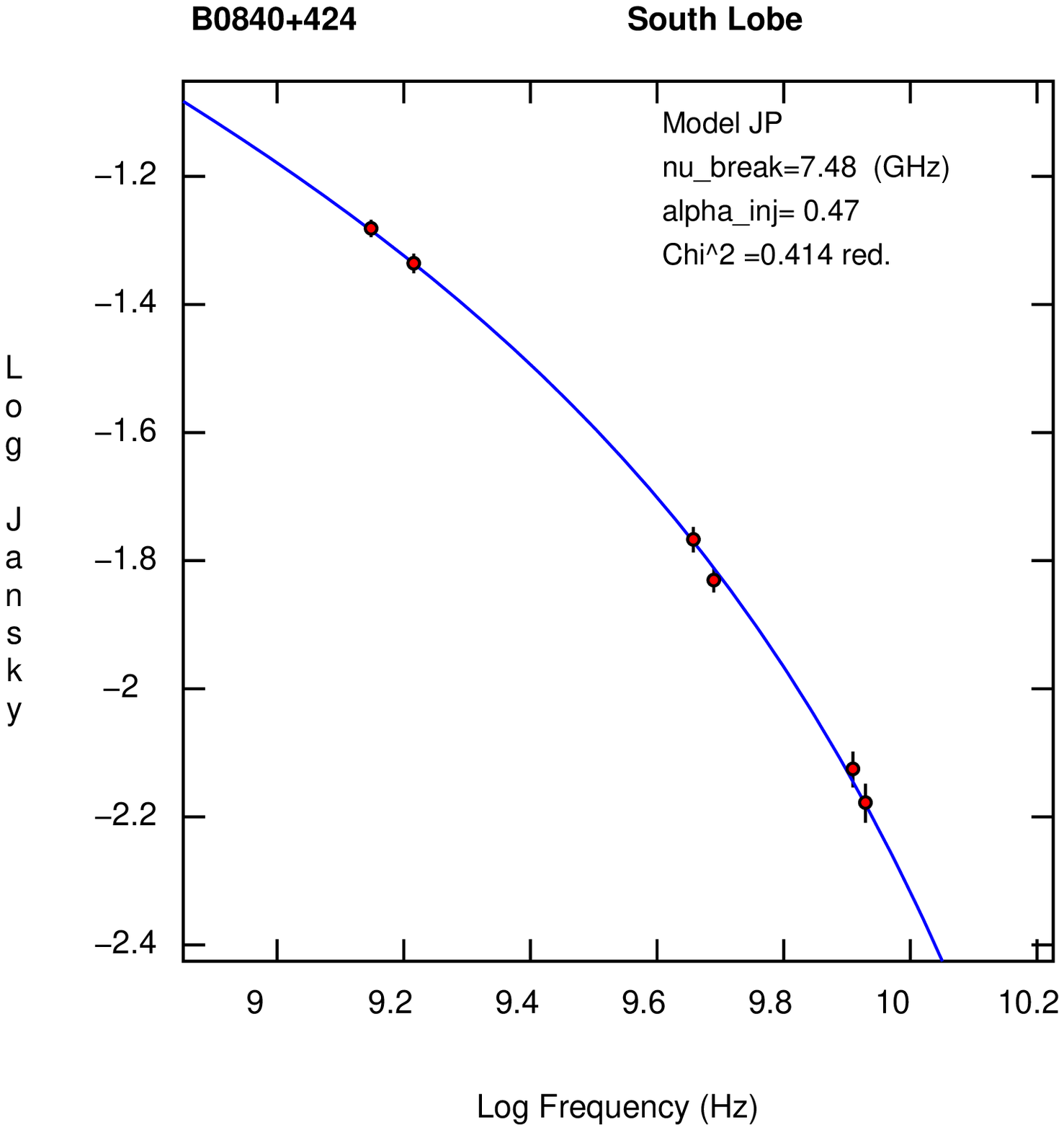}
\vspace{20.5cm}
\caption{The spectral fits in the lobes and hot-spots of B0147+400 and
  B0840+424. On each fit we report the model, the injection spectral
  index, the break frequency and reduced-chi-squared. The {\it upper
  panel} shows, clockwise, 
the local spectra in the Northern lobe (NL), Northern
  hot-spot (NH) and Southern hot-spot (SH) of the source
  B0147+400. The {\it bottom panel} shows, clockwise, 
the local spectra in the
  Norther hot-spot (NH), Southern hot-spot (SH) and Southern lobe (SL)
  of the source B0840+424. The spectra of the core components are not shown, since they can be found in Orienti et al. (\cite{mo04}).
The restoring beam (HPBW) is 8.20 $\times$
  4.46 mas in pa -3.59 and 1.198 $\times$ 0.753 in pa -1.94 for
  B0147+400 and B0848+424 respectively.}
\label{synage}
\end{center}
\end{figure*}

The radio source B0147+400 displays two rather compact features that
can be interpreted as hot-spots (labeled C in
  Fig. \ref{synage}), 
located $\sim$ 9 mas South-East and
$\sim$ 55 mas North-West of the core, with a flux density ratio of
S$_{SE}$/S$_{NW}$ $\sim$ 2.1 and 5.2 at C and X bands 
(Orienti et al. \cite{mo04}).\\
The brightest hot-spot (labeled SH in Fig. \ref{synage})
is embedded in the SE component and, with its
flux density (208, 158 and 121 mJy at L, C and X band
respectively, Dallacasa et al. \cite{dd02}; Orienti et
al. \cite{mo04}), dominates the radio emission ($\sim$ 68\%) in C
and X bands. 
It is best fitted by  
a power-law with an {\it injection} spectral index of $\alpha_{\rm
inj}$ $\sim$ 0.40.   
This implies that there is a continuous supply and re-acceleration of
fresh relativistic electrons, in agreement with what is predicted by the
source growth models.\\  
On the other hand, the faint hot-spot embedded in the NW component
(labeled NH) is well fitted by a higher injection spectral index of 
$\alpha_{\rm inj}$ $\sim$ 0.65.\\ 
In the source B0840+424 the Northern and Southern components are
$\sim$15 and $\sim$100 mas apart from the core (labeled C in
  Fig. \ref{synage}), with a flux density
ratio of S$_{N}$/S$_{S}$ $\sim$ 6.0 and 7.1 at C and X bands
(Orienti et al. \cite{mo04}).\\
The brightest hot-spot (labeled NH) is located at the centre of the
Northern component. The hot-spot can only be fitted by a
single-injection model, with an $\alpha_{\rm inj}$ of $\sim$ 0.40,
similarly to what is found in the bright hot-spot of
B0147+400. However,
this can be easily explained by considering a strong contamination by
the lobe emission. 
The hot-spot embedded in the Southern component (labeled SH) is not
bright and well-defined as in the Northern lobe, and it displays a
higher injection spectral index of $\alpha_{\rm inj}$ $\sim$ 0.50.\\
In both sources, the faintest hot-spots have $\alpha_{\rm inj}$
steeper than what is found by Orienti et al. (\cite{mo04}). 
However, that work, based on 3 independent frequencies only, provided
good fits for a range of injection spectral indices. The availability
of 6 independent spectral points in this study allows 
us to set stronger constraints
on the fits and their parameters.\\   

\subsection{The lobes}
The lobes represent the ideal loci where the radiative age can be
computed with a high degree of accuracy.
To estimate the source age, we determine the variation of the
break frequency across different regions of the extended components,
such as the lobes. To improve the reliability of the analysis, we
consider only those regions with a good signal-to-noise ratio at
all frequencies. Indeed, at high frequencies (i.e. X band), low surface
brightness regions are almost completely resolved out, causing the 
fits to fail.\\
Since the electrons in the lobes have likely received the last
acceleration in the corresponding hot-spot, we fit their spectra
with a single-injection model 
with the $\alpha_{\rm inj}$ derived for the hot-spot.\\
We try to fit the spectra with both the JP and KP models. The JP model
assumes that the {\it pitch angle $\theta_{p}$} between the electron
velocity and the magnetic field direction is continuously
re-isotropized, making the electrons age in the same way. In the KP
model, the {\it pitch angles} of the electrons populations are
constant, making electron populations with different $\theta_{p}$ age
in different ways.\\ 
Although both models provide similar fits to our spectra, the
  fits with the JP model have 
smaller reduced $\chi^{2}$ than the KP.\\ 
In B0147+400, we studied the break frequency across the Northern lobe
(labeled NL). Using the injection spectral index of $\alpha_{\rm inj}
= 0.65$ as derived for the northern hot-spot, we find
a minimum break frequency $\nu_{\rm br}$ = 14 GHz.\\
In the case of B0840+424, we fitted the spectra across the Southern
lobe (labeled SL) with a JP model with $\alpha_{\rm inj}$ of 0.50,
obtaining the lowest break frequency of 7.4 GHz.\\
For both sources, we have considered different regions across the
lobe, and the final choice was made on the best compromise between the
largest distance from the hot-spot and the need for significant
emission at the highest frequencies.

\subsection{The cores}
In both B0147+400 and B0840+400 the core, labeled C in
Fig. \ref{synage}, has been definitely identified by means of the
inverted spectrum displayed ($\alpha$ $\sim$ -0.5 and -0.03 for
B0147+400 and B0840+424, 
respectively; Orienti et al. \cite{mo04}).\\ 
These compact components are better visible
in the full resolution images at the highest observing frequency.
Indeed, in the X band they account for almost 10\% of the total 
flux density of the whole source. \\
Neither core shows any significant flux density variability in the 
X band in our data.\\
\begin{table*}
\begin{center}
\begin{tabular}{ccccccccccc}
\hline
\hline
&&&&&&&&&&\\
Source&z&LLS&B$_{eq}$&t$_{\rm syn}$&LS&Dist$_{\rm
  fhs}$&v$_{fhs}$&Age&Dist$_{\rm bhs}$&v$_{bhs}$\\
 & &pc&mG&10$^{3}$yr&pc&pc&c&10$^{3}$yr&pc&c\\
(1)&(2)&(3)&(4)&(5)&(6)&(7)&(8)&(9)&(10)&(11)\\
&&&&&&&&&&\\
B0147+400&0.20&210&3.0&  2.6& 64&188& 0.08&7.5&29&0.01\\ 
         &0.35&316&3.0&  2.6& 97&258& 0.13&7.5&44&0.02\\
         &0.40&344&3.0&  2.6&106&307& 0.13&7.5&48&0.02\\
         &0.60&429&3.3&  2.2&132&383& 0.19&6.5&60&0.03\\
         &0.80&488&3.6&  2.0&149&432& 0.25&5.6&67&0.04\\
         &1.00&518&4.0&  1.7&159&462& 0.31&4.8&72&0.05\\
B0840+424&0.20&362&1.2& 14.0&64&301&0.015&70.0&49&0.002\\
         &{\bf 0.35}&{\bf 544}&{\bf 1.6}&{\bf 9.4}&{\bf 97}&{\bf 452}&{\bf
  0.034}&{\bf 44.0}&{\bf 73}&{\bf 0.005}\\
         &0.40&592&1.6& 9.1&106&492&0.038&42.0&79&0.006\\
         &0.60&740&1.7& 8.3&132&614&0.052&38.0&100&0.009\\
         &0.80&834&1.9& 7.0&149&692&0.069&32.5&112&0.011\\
         &1.00&892&2.1& 6.0&159&740&0.086&28.0&120&0.014\\
&&&&&&&&&\\
\hline
\end{tabular} 
\vspace{0.5cm}
\end{center}   
\caption{Source ages. Column 1: the source name; Column 2: the
  redshift; Column 3: the total linear size;
Column 4: the equipartition magnetic field, computed
  assuming the source parameters from Orienti et al. (\cite{mo04});
Column 5: the radiative age, computed using equation \ref{equation1} and the
  lowest break frequency as described in Section 4.2; Column 6:
  the projected linear size, in pc, between the region considered to derive
  radiative age and the location of the corresponding hot-spot; 
Column 7: the projected distance, in pc, between the
  core component and the faintest hot-spot; Column 8: the mean
  advance speed of the faintest hot-spot;
Column 9: the source age;
  Column 10: the projected distance between the core and the brightest
  hot-spot; Column 11: the mean advance speed of the brightest hot-spot.} 
\label{age}
\end{table*}

\subsection{Radiative ages and the nature of CSOs sources}
From equation \ref{equation1}, it is clear that the synchrotron age is
strictly related to the break frequency $\nu_{\rm br}$, which can be
derived from the fits to the observed radio spectrum, and the
magnetic field. 
Direct measurements of the magnetic field are very difficult, often
impossible to carry out. Ideally, it can be measured from the turnover
frequency and component sizes, if both are known, but 
the uncertainties remain quite large. Alternatively, we can measure it 
by comparing
synchrotron and inverse Compton losses, but X-ray
observations of small and young radio sources have not provided strong
constraints so far.\\  
In this paper, the magnetic field of the source components has been
computed assuming minimum energy and equipartition conditions, and 
using standard formulae
(Pacholczyk \cite{pacho70}). Furthermore, proton and electron
energies have been assumed to be equal, with a filling factor of unity
(i.e. the source volume is fully and homogeneously filled by
relativistic plasma); an ellipsoidal geometry and an average optically
thin spectral index of 0.7 have been adopted. 
The observational parameters involved, such as the flux density and
the projected linear size of the source components, are from Orienti
et al. (\cite{mo04}).\\ 
We also point out that the magnetic field of CSS sources is of
a few orders of magnitude higher than that 'equivalent' to the Cosmic
Microwave Background radiation photons. Inverse Compton losses are,
therefore, neglected, the synchrotron being the main cooling mechanism.\\ 
Furthermore, even adiabatic losses are negligible, since the energy
spent due to the adiabatic expansion is three orders of
magnitude smaller than the synchrotron emission.\\ 
Unfortunately, both sources lack spectroscopic redshifts. The
source B0147+400 has no optical identification, while
B0840+400 can be identified with a galaxy in the 
Sloan Digital Sky Survey (SDSS). 
With the magnitudes provided by the SDSS for the source B0840+424, we
made use of the HyperZ code (Bolzonella et al. \cite{mb00}) to infer
the photometric redshift.
We obtained a photometric redshift of 0.35, with a probability of
83\%.\\
Since most of the intrinsic physical parameters, such as the magnetic
field (B$_{eq}$) and the linear size (LS), critically depend on
the redshift, for each source we provide a set of values considering
a few cases in which z is in the range of 0.2 - 1.0 (Table \ref{age}).
The radiative ages found in this way are in the range of 10$^{3}$ -
10$^{4}$ years.  
However, as previously mentioned, this should not be considered as the
source age.  
Models of source evolution predict that relativistic electrons are
deposited in the region of last acceleration, where they age, while
the hot-spot continues through the interstellar medium. 
Therefore, the electrons considered for the previous computation have no
memory of the original injection by the core, since they have
already been re-accelerated by the hot-spot.
The radiative ages
derived give us an indication of the time elapsed since the last
acceleration in the hot-spot.\\
Measuring the distance between the region considered for the radiative
age and the hot-spot, we can estimate the hot-spot advance speed. In
both sources, the hot-spot has covered a distance of $\sim$ 20 mas
from the region where we could measure the radiative age, which
implies a range of {\it mean} hot-spot velocities of 0.08c to
0.31c for B0147+400 and between 0.038c and 0.086c for
B0840+424,
depending on the source redshift (Table \ref{age}). 
Since we know the distance between the hot-spot and the core
component, and assuming a mean hot-spot velocity over the whole
  lifetime of the source (Polatidis \& Conway
\cite{pc03}), we can constrain the true source age. In the
case of B0147+400, the core-hot-spot separation is 53 mas, which leads
to a source age in the range of 4.8$\times$ 10$^{3}$ and 7.5$\times$
10$^{3}$. For the source B0840+424, the distance between the core and
the hot-spot is larger (97 mas), and we derive ages between
2.8$\times$ 10$^{4}$ and 7.0$\times$ 10$^{4}$ years, (4.4$\times$
10$^{4}$ years, considering the photometric redshift of 0.35).\\   
We can also estimate the mean advance speed of the brightest hot-spot,
dividing its distance from the core by the source age.
We obtain velocities ranging from 0.01c to 0.05c and 0.002c and 0.014c
for the brightest hot-spots of B0147+400 and
B0840+424 respectively (Table \ref{age}). 
With such velocities, it is not possible to detect any hot-spot advance
over a short period of time. We compared the data at the two epochs
available, observed about three years apart, by means of the MODELFIT and
the JMFIT, but we did not find any significant variation in the
position of the most compact source components. The accuracy in the
determination of the component position is in agreement with the
above results.

\subsection{Linearly Polarized Emission}
Images in the U and Q Stokes' parameters have been derived for both
the target sources, as well as for the calibration objects in C and X
bands.
Calibration sources proved to have integrated VLBA polarized emission
in agreement with VLA measurement carried out at a very near epoch,
as available from the VLA/VLBA polarization calibration database.\\
No significant ($\geq$ 3$\sigma$ noise level) polarized emission 
was detected for the target sources in C and X bands,
consistent with previous VLA observations (Fanti et al. \cite{cf01})
at the same frequencies, where both sources appear unresolved and
unpolarized at a resolution of $\sim$ 0''.4.\\ 
Polarization images are not shown.\\     
The local upper limits we could infer are of 0.03\% and 0.14\% for the Southern
and Northern components of B0147+400 respectively, and of 0.01\% and
0.08\% for the Northern and Southern components of B0840+424.\\ 
Our results are in good agreement with those found by Fanti et
al. (\cite{cf04}). 
In fact, although we do not know the redshift of both sources, we can
estimate their projected linear sizes $\leq$ 520 and 900 pc for
B0147+400 and B0840+424, respectively (Table \ref{age}). Following the
work of Fanti et al. (\cite{cf04}), both sources fall in the interval
of unpolarized CSS sources even in the X band, suggesting that the source
size is not large enough to lead the radio emission to emerge out of
the ``Faraday screen''.\\

 \begin{table}
\begin{center}
\begin{tabular}{cccccc}
\hline
\hline
&&&&&\\
Source&Component&S$_{\rm 8.4}$&$\theta_{\rm maj}$&$\theta_{\rm
  min}$&p$_{\rm min}$\\
 & &mJy&mas&mas&dyne/cm$^{2}$\\
&&&&&\\
B0147+400&HS&119&3&2&3.4$\times$10$^{-6}$\\
         &HN&22&9&6&4.9$\times$10$^{-7}$\\
B0840+424&HN&249&8&3&2.1$\times$10$^{-6}$\\
         &HS&26&13&8&1.4$\times$10$^{-7}$\\ 
&&&&&\\
\hline
\end{tabular} 
\vspace{0.5cm}
\end{center}   
\caption{The hot-spot pressure. Column 1: the source name; Column 2:
  the source component; Column 4: VLBA flux density at 8.4 GHz; Columns 5, 6: deconvolved angular size of the
  major and minor axis of the source component; Column 7: the minimum
  pressure, computed assuming the minimum energy condition, and an
  indicative redshift of 0.35.}  
\label{pressure}
\end{table}

\section{Discussion}
From the analysis of the synchrotron spectra in two CSOs, we
estimate radiative ages of about 5$\times$10$^{3}$ and
5$\times$10$^{4}$ years, in good agreement with kinematic and
radiative studies carried out on the same class of objects (Polatidis
\& Conway \cite{pc03}; Murgia et al. \cite{mm99}). 
Our measurements are not aimed to find
the ``accurate'' source age (which may be revised as new measurements
are added), but rather to determine whether the target sources are to
be considered {\it young} radio galaxies. \\
The approach used to infer the source age, described in Section 4.3, is
based on the strong assumption that the hot-spot velocity derived is
truly representative of the {\it mean individual hot-spot advance
speed}. However, there are several mechanisms that would cause the 
{\it instantaneous} hot-spot speed to vary, such as hydrodynamically
introduced internal pressure changes (Norman \cite{norman96}), 
as well as an inhomogeneous
external medium.\\
The large asymmetries both in arm-ratio and brightness shown by
B0147+400 and B0840+424 
strongly suggest that at least one of the two
aforementioned possibilities applies.
In Table \ref{pressure}, we report the hot-spot internal pressures
computed assuming that the source components are in the minimum energy
condition. 
Contrary to expectations, we find that in both sources the
hot-spot with the highest pressure is the slowest one,  
suggesting that the asymmetric morphology is more likely due to
an inhomogeneous clumpy ambient medium.\\
Using simple one-dimensional ram-pressure arguments, the advance speed
$v$ of the hot-spot is determined by the equilibrium between the
internal pressure $p_{i}$ and the ram-pressure of the external
medium:\\ 

\begin{equation}
p_{i} \propto n_{ext} m_{p} v^{2}
\label{equation2}
\end{equation}

where $n_{ext}$ is the particle density of the external medium and
$m_{p}$ is the proton mass. We assume an external density profile of
the King type, as suggested by X-ray observations of early-type galaxies 
(i.e. Trinchieri et al. \cite{tr86}). 
Since the total linear sizes of both sources are
smaller than the core radius ($<$ 1 kpc), we can assume a roughly
constant external gas density.\\
If in equation \ref{equation2} we consider
the average hot-spot velocity (Table \ref{age}) and the minimum
pressure (Table \ref{pressure}), computed at an indicative redshift of
0.35, for both sources we find that the brightest and closest hot-spot
is likely digging its way through a quite dense medium ($n_{ext}$
$\sim$ 5.7 and 50.0 cm$^{-3}$, for B0147+400 and B0840+424
respectively), similar to what one can expect in a cloud, while the
farthest component is likely moving through an intercloud medium
($n_{ext}$ $\sim$ 0.02 and 0.08 cm$^{-3}$, for 
B0147+400 and B0840+424 respectively) 
where the external density is
about 3 orders of magnitude smaller.\\  
Such clouds, indicating the presence of a rich and clumpy interstellar 
medium interacting with a CSS/GPS radio source, have been found by
means of high-resolution spectral studies of the neutral hydrogen (Morganti et
al. \cite{rm04}; Labiano et al. \cite{al06}).\\ 
In the GPS ULIRG galaxy 4C\,12.50, Morganti et al. (\cite{rm04})
detected a cloud with an \HI\ mass of a few 10$^{5}$ to 10$^{6}$
M$_{\odot}$ and $\sim$ 20$\times$66 pc in size, corresponding to a
density of $\sim$ 2$\times$10$^{3}$ cm$^{-3}$.\\
Labiano et al. (\cite{al06}) studied the \HI\ absorption in the two
very asymmetric CSS sources 3C\,49 and 3C\,268.3, in which the
brightest lobe is also the closest to the core, as in our targets.\\
In both sources, \HI\ absorption was detected in the brightest
(and closest) lobe only. 
The absorber medium has been interpreted in terms of clouds which are
in the environment of the GPS/CSS radio source, with densities of
220/$c_{f}$ cm$^{-3}$ (0.04$<$ $c_{f}$ $<$ 1, for 3C\,49)  and
360/$c_{f}$ 
cm$^{-3}$ (0.025$<$ $c_{f}$ $<$ 1, for 3C\,268.3), where
$c_{f}$ is the covering factor.\\     
These values are in good agreement with the characteristics of the ISM
in the Narrow Line Region (NLR; Fanti et al. \cite{cf95}), 
in which the hot-spots of the aforementioned sources actually reside.\\
Our results on B0147+400 and B0840+424 
are consistent with a picture
in which one side of the radio source is strongly interacting with a
dense cloud, while the other is expanding through an intercloud
medium. 
The interaction with the cloud causes the lobe to propagate more slowly, 
and favours radio emission by means of compression and shocks
(Jeyakumar et al. \cite{jeya05}; Bicknell et al. \cite{gb03}), 
which increase the energy production
efficiency. Furthermore, the clouds can also act as a Faraday screen,
leading to the observed depolarization as found in both sources.\\
The detection of such asymmetric
CSOs may be favoured by a selection effect. 
The interaction
with a dense ambient medium may enhance the radio emission, making these
objects more detectable.
This is in agreement with other studies based on the asymmetries of
CSOs (Saikia
et al. \cite{saikia03}), in which on such a small scale, the
probability that the brightest component is also the closest one to
the core is higher than in larger sources. 

\section{Conclusion}
We have presented the results of a new spectral analysis based on 
multi-frequency VLBA observations for two Compact Symmetric Objects
from the B3-VLA-CSS sample (Fanti et al. \cite{cf01}). The radiative
ages derived from the analysis of the break frequency are of about
5$\times$10$^{3}$ and 5$\times$10$^{4}$ years, supporting the
hypothesis that these are young objects.  
The individual hot-spot advance velocities range from 0.03c to 0.3c
for the farthest lobe, and from 0.005 to 0.05 for the closest one.\\ 
The strong asymmetries in the arm ratio and brightness found in 
both sources are more likely due to a strong influence exerted by a
clumpy and inhomogeneous medium, rather than a change in the hot-spot
internal pressure.  
The brightest and closest component is partially and temporarily
confined by a dense cloud which slows its propagation, while
the other component is expanding through a more diluted ambient
medium.\\      
From this result, we infer that the knowledge of the properties of the
ambient medium surrounding the radio source is of fundamental
importance in order to draw a complete and reliable picture of the
individual source evolution.\\  

\begin{acknowledgements}
We like to thank the anonymous referee for carefully reading the
manuscript and valuable suggestions.
The VLBA is operated by the US National Radio Astronomy Observatory
which is a facility of the National Science Foundation operated under
a cooperative agreement by Associated Universities, Inc. This work has
made use of the NASA/IPAC Extragalactic Database NED which is operated
by the JPL, California Institute of Technology, under contract with
the National Aeronautics and Space Administration. 
\end{acknowledgements}


\begin{thebibliography}{}

\bibitem[1990]{baum90}
Baum, S.A., O'Dea, C.P., de Bruyn, A.G., Murphy, D.W. 1990, A\&A, 232, 19 

\bibitem[2003]{gb03}
Bicknell, G.v., Saxton, C.J., Sutherland, R.S. 2003, PASA, 20, 102

\bibitem[2000]{mb00}
Bolzonella, M., Miralles, J.-M., Pell\'{o}, R. 2000, A\&A, 363, 476

\bibitem[2002]{dd02}
Dallacasa, D., Tinti, S., Fanti, C. et al. 2002, A\&A, 389, 115

\bibitem[1995]{cf95}
Fanti, C., Fanti, R., Dallacasa D., Schilizzi, R.T. et al. 1995, A\&A,
302, 317

\bibitem[2000]{cf00}
Fanti, C., Pozzi, F., Fanti, R. et al. 2000, A\&A, 358, 499

\bibitem[2001]{cf01}
Fanti, C., Pozzi, F., Dallacasa, D. et al. 2001, A\&A, 369, 380

\bibitem[2004]{cf04}
Fanti, C., Branchesi, M., Cotton, W.D. et al. 2004, A\&A, 427, 465

\bibitem[2005]{gu05}
Gugliucci, N.E., Taylor, G.B., Peck, A.B., Giroletti, M. 2005, ApJ,
622, 136

\bibitem[1974]{jp74}
Jaffe, W.J., Perola, G.C. 1974, A\&A, 26, 463

\bibitem[2005]{jeya05}
Jeyakumar, S., Wiita, P.J., Saikia, D.J., Hooda, J.S. 2005, A\&A, 432, 823

\bibitem[1962]{karda62}
Kardashev, N.S. 1962, SvA, 6, 317

\bibitem[2006]{al06}
Labiano, A., Vermeulen, R.C., Barthel, C.P. et al. 2006, A\&A, 447, 481

\bibitem[2004]{rm04}
Morganti, R., Oosterloo, T.A., Tadhunter, C.N. et al. 2004, A\&A, 424, 119

\bibitem[1999]{mm99}
Murgia, M., Fanti, C., Fanti, R. et al. 1999, A\&A, 345, 769

\bibitem[2000]{mmphd}
Murgia, M. 2000, PhD Thesis

\bibitem[2003]{mm03}
Murgia, M. 2003, PASA, 20, 19

\bibitem[1996]{norman96}
Norman, M. 1996, ASPC, 100, 405

\bibitem[2004]{mo04}
Orienti, M., Dallacasa, D., Fanti, C. et al. 2004, A\&A, 426, 463

\bibitem[1970]{pacho70}
Pacholkczyk, A.G. 1970, Radio Astrophysics (San Francisco: Freeman \& Co.)

\bibitem[2003]{pc03}
Polatidis, A.G., Conway, J.E. 2003, PASA, 20, 69

\bibitem[1996]{rh96}
Readhead, A.C.S., Taylor, G.B., Xu, W. et al. 1996, ApJ, 460, 612

\bibitem[2003]{saikia03}
Saikia, D.J., Jeyakumar, S., Mantovani, F. et al. 2003, PASA, 20, 50

\bibitem[2005]{sa05}
Siemiginowska, A., Cheung, C.C., LaMassa, S. et al. 2005, ApJ, 632, 110 

\bibitem[2000]{sn00}
Snellen, I.A.G., Schilizzi, R.T., Miley, G.K. et al. 2000, MNRAS, 319, 445

\bibitem[1986]{tr86}
Trinchieri, G., Fabbiano, G., Canizares, C.R. 1986, AJ, 310, 637

\bibitem[1984]{breu84}
van Breugel, W., Miley, G., Heckman, T. 1984, AJ, 89, 5

\bibitem[1994]{wil94}
Wilkinson, P.N., Polatidis, A.G., Readhead, A.C.S. et al. 1994, ApJ,
432, 87

\end{thebibliography}
\end{document}